\documentclass{iopart}

\usepackage{amssymb}
\usepackage[dvips]{graphicx}
\usepackage{mathrsfs}

\newcommand{\be}{\begin{equation}}
\newcommand{\ee}{\end{equation}}
\newcommand{\bel}[1]{\begin{equation}\label{#1}}
\newcommand{\ba}{\begin{eqnarray}}
\newcommand{\ea}{\end{eqnarray}}
\newcommand{\bal}[1]{\begin{eqnarray}\label{#1}}

\begin{document}

\title{A Three-Stage Search for Supermassive Black Hole Binaries in LISA Data}

\date{\today}

\author{Duncan A Brown$^{1,2}$, Jeff Crowder$^{3}$, Curt Cutler$^{2,3}$, Ilya
Mandel$^{2}$ and Michele Vallisneri$^{2,3}$}
\address{$^1$LIGO Laboratory, California Institute of Technology, 
Pasadena, CA 91125}
\address{$^2$Theoretical Astrophysics, California Institute of Technology, 
Pasadena, CA 91125}
\address{$^3$Jet Propulsion Laboratory, California Institute of Technology, 
Pasadena, CA 91109}

\begin{abstract}
Gravitational waves from the inspiral and coalescence of supermassive
black-hole (SMBH) binaries with masses $m_1 \sim m_2 \sim 10^6$ are
likely to be one of the strongest sources for the Laser Interferometer
Space Antenna (LISA). We describe a three-stage data-analysis pipeline
designed to search for and measure the parameters of SMBH binaries in
LISA data.  The first stage uses a time--frequency track-search method
to search for inspiral signals and provide a coarse estimate of the
black-hole masses $m_1, m_2$ and of the coalescence time of the binary
$t_c$.  The second stage uses a sequence of matched-filter template
banks, seeded by the first stage, to improve the measurement accuracy of
the masses and coalescence time. Finally, a Markov Chain Monte Carlo
search is used to estimate all nine physical parameters of the binary
(masses, coalescence time, distance, initial phase, sky position and
orientation). Using results from the second stage substantially shortens
the Markov Chain burn-in time and allows us to determine the number of
SMBH-binary signals in the data before starting parameter estimation. We
demonstrate our analysis pipeline using simulated data from the first
LISA Mock Data Challenge.  We discuss our plan for improving this pipeline 
and the challenges that will be faced in real
LISA data analysis.
\end{abstract}

\maketitle

\section{Introduction} 

There is compelling evidence from electromagnetic observations that the cores
of galaxies contain supermassive black holes (SMBHs)~\cite{Krolik:1999}. SMBH
binaries can form after galactic mergers as the black holes from the
individual galaxies fall to the center of the merged system and form a bound
pair.  Hierarchical-merger models of galaxy formation predict that SMBH
binaries will be common in
galaxies~\cite{2002MNRAS.336L..61H,2003ApJ...582..559V} and the presence of one such binary
has been inferred from X-ray measurements of the core of the galaxy
NCG~6240~\cite{Komossa:2002tn}.  The evolution of an SMBH binary will
eventually be driven by radiation reaction from the emission of
gravitational waves (GWs) and the binary will inspiral and merge to form a single SMBH. The
GWs from inspirals of SMBH binaries with component masses $m$ in the 
range
$m \sim 10^4$--$10^7\,M_\odot$ will be one of the strongest sources for LISA,
the planned space-based GW
detector~\cite{YellowBook,Haehnelt:1998hd}. The direct detection of SMBH
binaries will be of wide astrophysical relevance, for example by probing the
merger rates and histories of galaxies~\cite{Hughes:2001ya}, or by providing
cosmological standard candles~\cite{Holz:2002cn}.

Searching for SMBH binary inspiral signals is expected be one of the 
more
straightforward tasks in LISA data analysis. The velocities of the black holes
during the inspiral are $v/c \ll 1$, and so existing post-Newtonian
waveforms~\cite{Arun:2004ff,Blanchet:2001ax} will describe the
gravitational waveforms with sufficient accuracy for use as templates in a
matched-filter search~\cite{Wainstein:1962}. As such, searches for SMBH
binaries in LISA data will be similar in nature to existing searches for
binary--neutron-star (BNS) inspirals in ground-based GW
detectors, such as the Laser Interferometer Gravitational-wave Observatory
(LIGO)~\cite{Barish:1999vh}.  However, there are several key differences between LIGO
and LISA binary inspiral searches.  First, the LIGO pipelines are
designed to search for signals with expected signal-to-noise ratios (SNRs) $\lesssim 10$, whereas the
SNR of LISA SMBH binaries at distances $z \lesssim 2$ is expected to
be several hundred or more.  Second, the BNS signals sweep through the
sensitive frequency band of ground-based detectors on timescales of order a
minute, during which detector velocities and orientations can be considered as
fixed to high accuracy. By contrast, LISA will be able to observe a single
SMBH inspiral for weeks to months.  During that time, the LISA velocity and
orientation change appreciably, inducing modulations in the recorded signal.
Indeed, almost all the information about an SMBH binary's sky location and
orientation is encoded in these modulations. (In the ground-based case, 
a
network of three or more widely separated detectors is required to determine a
binary's sky location by triangulation between the times of arrival of the GW
signals at the different detector locations.) Finally, whereas the rate 
of BNS
inspirals in ground-based detectors makes it unlikely that multiple signals
will be observed concurrently, LISA data may contain simultaneous 
signals from a few different SMBH binaries.

Existing search pipelines developed for ground-based observations of
stellar-mass binary inspirals can achieve high detection efficiency already at
SNRs $\sim 10$~\cite{Babak:2006ty,Allen:2005fk,Brown:2004pv,Abbott:2006zx}, so
the task of detecting SMBH inspirals with LISA seems easy in comparison.
Furthermore, since SMBH binaries at $z \sim 1$ have such high SNR, and because
of LISA's relatively wider frequency band (roughly three orders of magnitude
for LISA, compared to two for LIGO), it should also be possible to determine
the masses and spins of the binaries with significantly higher accuracy in the
LISA case than for ground-based detections.  Fisher-matrix
calculations suggest that, for SMBHs detected at $z \sim 1$, LISA should be able to determine the chirp mass to relative accuracy $\sim 10^{-5}$, both individual masses to $\sim 10^{-3}$
and the SMBH spins to $\sim 10^{-3}\mbox{--}10^{-2}$~\cite{Lang:2006bz}. Indeed, the
goal of our data-analysis pipeline is not only to detect the SMBH signals, but
also to provide accurate measurements of the binary parameters.

Based partly on the considerations discussed above, our group has adopted the
following three-stage search method.  Low-$z$ SMBH binary inspirals are so bright
that they are  easily visible as tracks in time--frequency (TF) 
spectrograms.
Therefore our first stage consists of a search for such TF tracks; the shape
and location of the track yields a first estimate of the two masses, $m_1$ and
$m_2$, and the coalescence time, $t_c$.  The second stage is a set of more
refined grid-based matched-filter searches that start in a neighborhood of
the best-fit parameters found in the first stage; these searches  home in on
more accurate values for the three parameters $m_1$, $m_2$ and $t_c$.  The
final stage is currently a straightforward  implementation of a Markov Chain
Monte Carlo (MCMC) simulated-annealing search for the best-fit parameters in
the full nine-dimensional parameter space (including also the binary's luminosity
distance, initial phase, inclination, polarization, ecliptic latitude and
longitude).

There are a few reasons for adopting such a complicated algorithm. First, we
believe that the capability of looking for TF tracks is a very useful one to
develop in the LISA context: it is possible that there will be tracks that do
{\it not} follow the expected chirping pattern, and so would not be found by
more sophisticated (grid-based or MCMC) methods, even though they are visible
to the eye. The track-search method also allows us to count the number of SMBH
binary signals present in the data before attempting parameter estimation.
Second, the grid search is useful to make sure that we do not miss any binary
sources, by examining the entire parameter space. In the pipeline described
here, however, we did not cover the entire parameter space in our grid search;
rather, we seeded the second-stage search using the parameters obtained from
the first stage. In future implementations, we intend to compare the full
grid search to this method. Finally, the MCMC approach is clearly very adept
at obtaining the final parameter estimates.

We have tested the performance of our SMBH binary search pipeline using data
from the Mock LISA Data Challenges (MLDCs)~\cite{Arnaud:2006gm,Arnaud:2006gn}.
The MLDCs are a program sponsored by the LISA International Science Team to
foster the development of LISA data-analysis methods and tools, and to
demonstrate already acquired milestones in the extraction of science
information from the LISA data output.  In the MLDCs, GW
signals whose parameter values are unknown to the challenge participants are
embedded in synthetic LISA noise; participants are challenged to identify the
signals and extract their parameters. Challenges of increasing difficulty are
being issued roughly every six months. The results from the first Challenge
are summarized by Arnaud and colleagues in this volume~\cite{Arnaud:2007vr}.
Challenge 1 included two datasets with signals from isolated
SMBH systems; we analyzed one of them. One of the goals of the MLDCs is
to demonstrate that data-analysis pipelines can actually achieve the
fantastic parameter measurement accuracy predicted by the Fisher-matrix
analysis.  

Two other differences between the ground-based and space-based cases deserve
mention. First, SMBH binaries may enter the LISA band with considerable
eccentricity, whereas the BNSs observed by ground-based
detectors will have become essentially circular by the time they enter the
observation band.  Second, in the ground-based case the binary-inspiral
signals are immersed in noise that originates almost entirely from the
instrument, while through much of LISA's sensitivity band the dominant noise
comes from unresolved Galactic white-dwarf binaries. To keep Challenge 1
relatively simple, however, these last two complications were omitted in
creating the synthetic datasets, and hence from our initial pipeline described
here.

The rest of this paper is organized as follows: in
sections~\ref{s:stage1}--\ref{s:stage3} we describe the three stages in
our SMBH binary data-analysis pipeline: a track search in the
time-frequency plane, a grid-based matched filtering search, and Markov
Chain Monte Carlo; in \ref{s:mldcresults} we present the results of
analyzing the MLDC dataset 1.2.1; and in \ref{s:future} we discuss our
plans for improving the pipeline to cope with issues such as binary
eccentricity and the noise sources likely to be observed in real LISA
data.

\section{Stage 1: Search for tracks in the time--frequency plane}
\label{s:stage1}

The TF spectrogram contains enough information to identify an SMBH binary inspiral at a high SNR. The techniques described below make it possible to quickly search for the presence of an SMBH binary inspiral in the signal and to get rough estimates of its parameters.

Challenge 1 includes signals from the adiabatic inspiral of a circular binary
system of nonspinning SMBHs. The frequency evolution of these inspirals is given
by (7.11a) of~\cite{Will:1996zj} in terms of the time of coalescence 
$t_c$ and the two SMBH masses $m_1$ and $m_2$. We write it here as a 
function of the symmetric mass ratio $\eta=m_1 m_2/(m_1+m_2)^2$ and the 
chirp mass $M_c=(m_1 + m_2) \eta^{3/5}$, using the second-order 
post-Newtonian (2PN) approximation:
\begin{eqnarray}
\fl
\label{WW}
f_\mathrm{GW}(t) = \frac{\eta^{3/5}}{8\pi M_c} (T_c-T)^{-3/8} 
	\Biggl\{1+ \biggl[\frac{743}{2688}+\frac{11}{32}\eta\biggr] 
		(T_c-T)^{-1/4} -\frac{3\pi}{10} (T_c-T)^{-3/8} \\
	+ \biggl[\frac{1855099}{14450688}+\frac{56975}{258048}\eta+
	\frac{371}{2048}\eta^2\biggr](T_c-T)^{-1/2} + 
	\mathcal{O}{\biggl[(T_c-T)^{-5/8}\biggr]}\Biggr\}. \nonumber
\end{eqnarray}
Here $f_\mathrm{GW}$ is the GW frequency in Hz, $M_c$ is expressed in seconds, and $T$ is the dimensionless time variable 
related to coordinate time $t$ by $T=t\,\eta^{8/5}/(5 M_c)$.

\begin{figure}
\includegraphics[width=\textwidth]{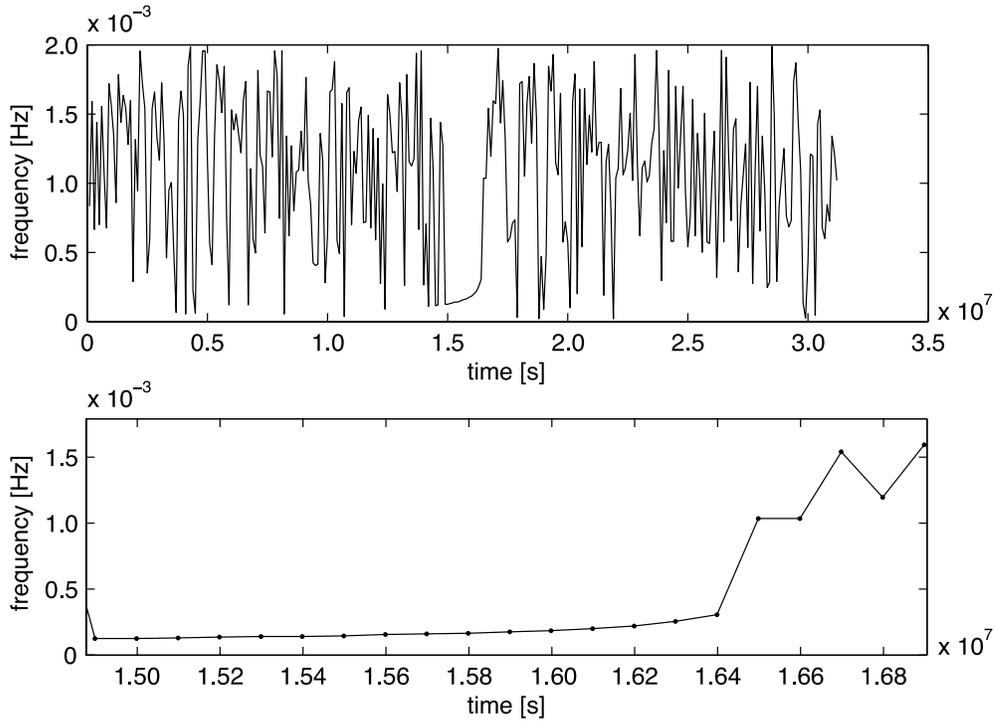}
\caption{Time--frequency plot of the brightest pixel in each time bin, as computed for the $X$ channel of Challenge 1 training set 1.2.1. The bottom plot is a blown-up version of the top plot showing the presumed track found on the first pass through the data.} 
\label{Fig:tf}
\end{figure}

We create a TF map of the noisy data stream $s(t)=h(t)+n(t)$ [in fact, one of the Time-Delay Interferometry (TDI) channels $X(t)$, $Y(t)$ and $Z(t)$ provided in the MLDC datasets], sampled with timestep $dt$, in two passes.  On the first
pass, we split up the data stream into time bins of equal duration $\Delta t$. The TF spectrogram will then consist of pixels of size  $\Delta t \times \Delta f$, where $\Delta f=1/(\Delta t)$. We determine the normalized power contained in each pixel with a Fast Fourier Transform (FFT), normalizing by the power spectral density of the noise, and then
find the peak frequency in each bin by searching for the loudest pixel (see
below for details).  The resulting set of $\{$time, frequency$\}$ pairs allows
us to search for an inspiral track on the TF map (see
figure \ref{Fig:tf}).  Once such a track is identified, we make a second pass through the data, iterating through the
track region with time bins of varying duration to create an improved
TF map. Earlier in the track, a larger $\Delta t$ helps to detect
a weak signal and achieve greater frequency resolution;
closer to coalescence, a smaller $\Delta t$ reduces the error in estimating the rapidly chirping GW frequency.

In fact, we have made several improvements to the general approach outlined in the previous paragraph.  The first set of improvements concerns the determination of the peak frequency in a given time bin.  Simply searching for
the loudest pixel would give frequency-determination errors of order $1/(\Delta f)$, even for a noiseless signal. Instead, we achieve higher accuracy by modeling the bleeding of frequency into neighboring pixels:   specifically, we determine the peak frequency by fitting the
logarithm of power in the pixels nearest to the brightest pixel to a parabola, using zero-padding in the time domain to achieve better
frequency resolution when necessary.  We also apply a Hanning window to
the signal prior to taking the FFT, and we overlap time bins to avoid 
information loss from windowing.  

Another improvement concerns the variable timestep and the selection of outliers on the second pass through the data. If the peak frequencies of neighboring time bins differ by more than $2\Delta f$, we decrease $\Delta t$ by a pre-set factor (say, $1.5$) to reduce the sweep of frequency in each bin. If this operation fails to bring the peak frequencies closer together, we declare the data point an outlier, and skip to the next bin.

\begin{figure}
\begin{indented}
\item[]\includegraphics[width=3.5in]{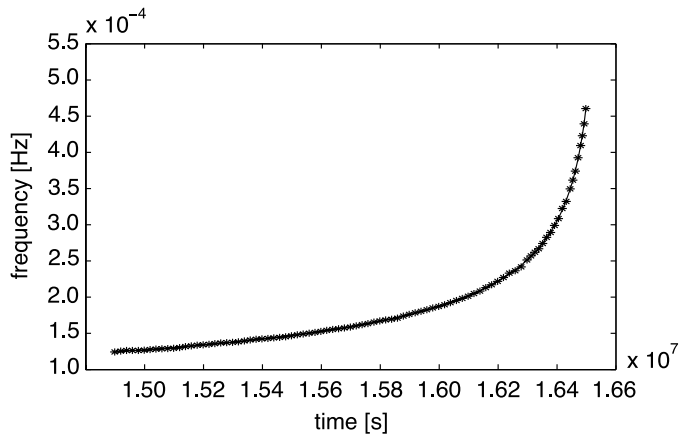}
\end{indented}
\caption{The stars represent individual points on the TF map
obtained during the second pass through the data in the $X(t)$ channel of where
Training Set 1.2.1. 
The curve is the result of fitting these points to the model (\ref{WW}).}
\label{Fig:fit}
\end{figure}

The $\{$time, frequency$\}$ data points obtained
on the second pass serve as inputs to a MATLAB least-squares fitting algorithm that extracts the inspiral parameters  $t_c$, $M_c$ and $\eta$  by fitting these
points to the model of (\ref{WW}) (see figure \ref{Fig:fit}). Specifically, we  find 
the values
$\hat{t}_c$, $\hat{M_c}$ and $\hat{\eta}$ that minimize the sum
\bel{ls}
\Sigma=\sum_{i=1}^N \left[f(t_i)-f_\mathrm{GW}(t_i;\, t_c, M_c,\eta)\right]^2,
\ee
where the $t_i$ are the centers of the output time bins, $f(t_i)$ are the
associated frequencies, and $f_\mathrm{GW}(t_i;\, t_c,M_c,\eta)$ is the model from (\ref{WW}).

Although one could weight the data points on the basis of the signal amplitude, such a weighting seems to carry little benefit: late in the inspiral, the increased amplitude offers greater SNR, which is however substantially offset by poorer frequency determination (due either to frequency drift within each time bin if $\Delta t$ is not properly adjusted, or to low frequency resolution if it is).

\begin{table}
\caption{\label{table:tfparams}Parameters extracted via TF searches from the $X$, $Y$ and $Z$ channels of 
blind Challenge dataset 1.2.1.  $N$ is the number of data points obtained during
the second pass through the data and $\Sigma$ is the sum of the squares of the residuals, as defined
in (\ref{ls}).}
\begin{indented}
\lineup
\item[]\begin{tabular}{@{}l@{\quad}l@{\quad}l@{\quad}l@{\quad}l@{\quad}l}
\br
 & $N$  & $\Sigma/10^{-11}$ & 
	$M_c/(10^6 M_\odot)$ & $\eta$ & $t_c/(10^7 \, {\rm s})$ \\
\mr
 $X$ & 156 & \09.2 & 1.2096 & 0.182 & 1.3373\\
 $Y$ & 190 & \09.21 & 1.2033 & 0.139 & 1.3370\\
 $Z$ & 192 & 11.5 & 1.2099 & 0.183 & 1.3373\\
\br
\end{tabular}
\end{indented}
\end{table}

Table~\ref{table:tfparams} shows the results of the TF search 
on the blind Challenge dataset 1.2.1. After averaging results from the three TDI streams, we found $\hat{M}_c=1.208\times 10^6\ M_\odot$, $\hat{\eta}=0.17$ and $\hat{t}_c=1.3372 \times 10^7$ s.   The accuracy of these estimates is discussed in Sec.~V below; suffice it to say that these first-stage results were certainly accurate enough for our purpose. 

\section{Stage 2: Grid-based search}
\label{s:stage2}

The grid-based part of the search relies on the template placement
algorithm of Babak~\emph{et.~al.}~\cite{Babak:2006ty} and the
\textsc{findchirp} matched filtering algorithm of
Allen~\emph{et.~al.}~\cite{Allen:2005fk}, both of which were developed
for the LIGO binary neutron star searches. The basic
algorithm is as follows: a grid of templates is constructed in the
$(m_1,m_2)$ plane using the metric-based square-grid placement
algorithm~\cite{Owen:1998dk,Babak:2006ty} implemented in the LIGO
Algorithm Library (LAL)~\cite{LAL}\footnote{Babak~\emph{et.~al.} also
describe a more efficient hexagonal placement algorithm, however we were
unable to place templates for LISA SMBH binaries using the LAL
implementation of this algorithm. We intend to work with the authors of
the LAL code to resolve this.}. The fineness of the grid is specified by
its minimum-match parameter $\mathrm{MM}$, which is the minimum overlap
between any point in the parameter space and its nearest grid-point.  To
implement the algorithms described in~\cite{Allen:2005fk}, we have
written C code which implements the matched filtering algorithms and
template generation. These C functions are then ``wrapped'' by the
Simplified Wrapper Interface Generator (SWIG), which allows them to be
called from the Python high-level programming language. This approach
allows us to rapidly prototype and develop the procedure described 
below.

For each mass pair in the grid,
we compute a (Fourier-transformed) waveform $\tilde h(f)$  (corresponding to
coalescence at $t=0$), using 2PN waveforms and
the stationary phase approximation (SPA)~\cite{Droz:1999qx}.  We transform
from $\tilde h(f)$ to the LISA TDI variable $\tilde X_h(f)$ using
\begin{equation}
\tilde X_h(f) = {\rm sin}^2(2\pi f L) \tilde h(f),
\end{equation}
where $L$ is the LISA arm length. Let the (Fourier transformed) data be
$\tilde X_s(f)$; then for each template waveform $\tilde X_h(f)$ in our grid
we use the FFT to compute the inverse Fourier transform
\begin{equation}
z(t) = \int \frac{\tilde{X}_s(f) \tilde{X}^*_h(f)}{S_X(f)} \rme^{2\pi i t f} \, df,  
\end{equation}
and we maximize $|z(t)|$ over $t$ to estimate the time of coalescence.  We
identify the best-fit point in the $(m_1,m_2)$ plane, and then repeat the
search in a neighborhood of that point, with a finer grid. We do this four
times, with a final minimum-match parameter $\mathrm{MM} = 0.995$.  
For Challenge 1.2.1, based on the results from the TF stage 
($m_1 \approx 2.9 \times 10^6 M_{\odot}$ and 
$m_2 \approx 7.3 \times 10^5 M_{\odot}$), we chose our initial
grid to cover the portion of the $(m_1,m_2)$ plane satisfying $6\times 10^5 < m_2 < m_1 < 3.2 \times 10^6 M_{\odot}$,  with initial $\mathrm{MM}  = 0.30$.

Now, our parameter-estimation errors are dominated not by the coarseness of
the grid, but by the fact that our 2PN SPA  waveforms are not identical to BBH
waveforms injected into the Mock LISA data, even for the same parameter
values. Our 2PN SPA waveforms differ from the MLDC versions by higher-order PN
terms, and do not include the modulations due to the detector motion. They are
also simply cut off at the frequency of the innermost stable circular orbit
(ISCO) of a test mass in the Schwarzschild spacetime, while the MLDC waveforms
end with a very particular choice of taper. Therefore we do one final grid
search using MLDC waveforms (again with $\mathrm{MM}=0.995$), and for some
particular choice of the five angles $(\theta, \phi, \iota, \psi, \varphi_0)$.
Although these angles are wrong, in this step the \emph{other} features of the
templates (e.g., the Doppler modulation of the frequency due to LISA's orbit
and the amplitude taper) do match those of the injected MLDC binary waveforms,
and so presumably yield improved parameter estimates.

\section{Stage 3: Markov Chain Monte Carlo}
\label{s:stage3}

So far, the first two stages have given estimates only of the two masses and coalescence time; in addition, the stage-2 analysis was based only on the $X$ channel.  Thus, we rely on the MCMC stage to find the distance, sky location, and the polarization and inclination angles of our source. A more efficient way to do this would be to use the 
${\mathcal F}$-statistic~\cite{Jaranowski:1998qm,2004PhRvD..70b2003K} to automatically optimize over four amplitude parameters that are functions of distance, polarization, inclination and initial phase; however, we did not have time to implement this procedure for Challenge 1. Therefore our MCMC code does a brute force search over all parameters---but with the advantage that it starts in the right vicinity for the masses and coalescence time, as estimated in the first two stages. 

MCMC approaches have shown promise in the extraction of GW-source parameters with LISA~\cite{Cornish:2005qw,Cornish:2006ry,Wickham:2006af,Cornish:2006dt,Stroeer:2006ye,Crowder:2006eu}. Nevertheless, it has been suggested that, for SMBH binaries, MCMC searches over a full parameter set need to be started in a neighborhood of the correct source parameters to efficiently characterize the posterior probability density functions~\cite{Cornish:2006ry}. Since the initial 
search grid provided a good estimate of three parameters (the constituent masses and coalescence time $t_c$), and since it is trivial to extremize 
analytically over the luminosity distance, we were hopeful that we could determine the values of the sky location and binary 
orientation with a straightforward implementation of the Metropolis--Hastings Algorithm (MHA). Since time was limited and posterior distributions were not required for Challenge 1, we chose not to estimate these, but rather to use the MHA to locate the best-fit parameters.

In the MHA, a Markov chain is built by accepting a new proposed point with probability $\alpha = {\rm min} (1,H)$; $H$ is the Hastings ratio for a 
jump from position $\vec{x}$ to $\vec{y}$ in parameter space, given by
\begin{equation}
H = \frac{p(\vec{y}) p(s \vert \vec{y}) q(\vec{x} \vert \vec{y})} {p(\vec{x}) p(s \vert \vec{x}) q(\vec{y} \vert \vec{x})} \, ,
\end{equation}
where $p(\vec{x})$ is the prior distribution, $p(s \vert \vec{y})$ is the likelihood of the parameter set $\vec{y}$ producing the signal $s 
$, and $ q(\vec{x} \vert \vec{y})$ is the proposal distribution used to generate the move from $\vec{x}$ to $\vec{y}$. If the noise is a normal 
process with zero mean, the likelihood is given by
\begin{equation}\label{likely}
p(s \vert \vec{\lambda}) \propto \exp - \bigl( s - h (\vec{\lambda}) \big| s - h(\vec{\lambda}) \bigr) / 2 \,,\end{equation}
with ``$(\cdot|\cdot)$'' the standard inner product computed with respect to the LISA instrument noise.

The Markov chain process is guaranteed to converge to the posterior probability distribution if the proposal distribution is nontrivial; however, the speed of convergence does depend on its choice. In this search we adopted two types of proposals: the first consisted of a multivariate normal distribution with jumps directed along the eigendirections of the Fisher information matrix, computed locally; the second amounted to drawing parameters from uniform distributions. For the angular parameters, both timid and bold draws (from small or large ranges)
were made to ensure we were fully exploring parameter space; for the component masses, only timid draws ($< 1\%$) were used.

Multiple concurrent chains were started using the parameter estimates obtained in stage 2.  These were run on a supercomputing cluster with 3.2 GHz Intel Pentium 4 processors, using Synthetic LISA~\cite{Vallisneri:2004bn} to reproduce the LISA response to the SMBH binary waveforms. Each run was limited to $12$ hours, providing $\sim 3,500$ steps in each of the chains. The most promising candidates at the end of the first run were used as the starting locations of a second run. At the end of the first run the best candidates had reached log
likelihood values in the neighborhood of $200,000$; the second run saw them increase to $\sim 205,000$. The chains converged around two points in parameter space, differing by their locations on opposite sides of the sky. This was not unexpected: dual maxima at antipodal sky positions are a well-known degeneracy for LISA sources. Our choice between the two final parameter sets was based on a visual comparison of the putative signals with the challenge dataset.

In future implementations of the pipeline, we plan to incorporate the ${\mathcal F}$-statistic in the MCMC stage to reduce the size of parameter space. This will increase search efficiency and relax the need to begin the search in a neighborhood of the best-fit parameters (something that will be necessary when searching for the dimmer SMBH binaries of Challenge 2). Another time-saving measure will be to start the search on a limited portion of the data stream, and then steadily increase its size. This process, called frequency annealing~\cite{Cornish:2007jv}, allows a quick initial exploration of parameter space, and a careful later investigation of the exquisitely sharp likelihood peaks close to bright SMBH binaries. 

\section{Results for MLDC Challenge}
\label{s:mldcresults}

     As was the case for many Challenge-1 participants, the Dec.\ 3,
2006 submission deadline arrived before our pipeline was fully ready;
nevertheless we decided to submit our best estimates for the parameters
of the blind dataset 1.2.1. This dataset consisted of the three TDI
unequal-Michelson channels $X(t)$, $Y(t)$ and $Z(t)$.  In stage 1 of our
search, we analyzed each of these channels separately, and simply
averaged the three results to arrive at the stage-one parameter
estimates shown in the fourth column of table~\ref{results}. In stage 2,
only the $X(t)$ data was analyzed (partly because of time pressure). In
stage 3, we analyzed two orthogonal TDI channels given by $X$ and $(X +
2Y)/\sqrt{3}$.
    
\begin{figure}[ht]
\includegraphics[width=\textwidth]{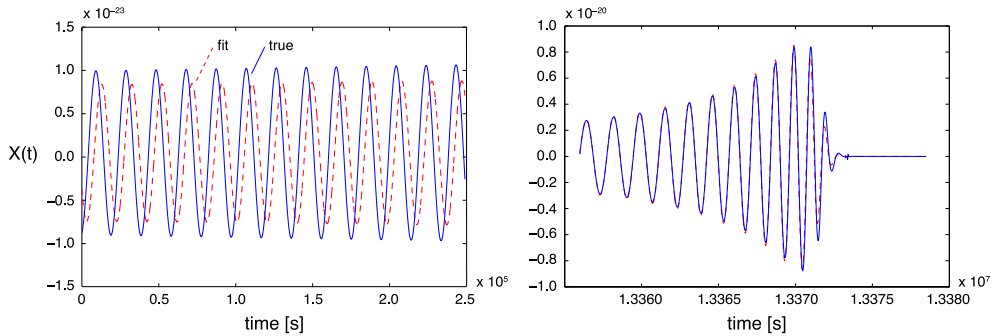}
\caption{Comparison of our best-fit $X(t)$ to the true $X(t)$ for a) a short stretch of time near $t_c$ and b) a short stretch near the beginning of the dataset.  Clearly, our fit is excellent near $t_c$, where most of the SNR accumulates, but much poorer at early times.}
\label{Fig:Xtfit}
\end{figure}

      The true signal parameters were made publicly available on Dec.\
4, and here we briefly describe how our search fared in their recovery.
The injected signal had a combined\footnote{In this context, $A$ and $E$
are the orthogonal, optimal TDI observables given by $(2X - Y - Z)/3$
and $(Z-Y)/\sqrt{3}$, as used in \cite{Arnaud:2007vr}.  The third
orthogonal, optimal TDI observable, $T$, contributes only a tiny
fraction of the total SNR for these sources.} $(A+E)$ SNR of $667.734$;  
its true physical parameters are listed in the third column of
table~\ref{results}. Our best-fit waveform matched the true waveform
rather well: it had an SNR of $664.47$ and its cross-correlation with
the true waveform was 0.994 for the $A$ channel and 0.996 for the $E$
channel~\cite{Arnaud:2007yj}. The quality of the fit is illustrated in
figure \ref{Fig:Xtfit}, which compares the true $X(t)$ (produced by us
from the key file) with our best-fit $X(t)$, for short time stretches
near the coalescence time $t_c$ and near the beginning of the dataset.
Clearly our fit is excellent near $t_c$, where most of the SNR
accumulates, but is much poorer at early times, when the contribution to
the SNR is much lower. The lesson from the other two Michelson variables
is qualitatively the same.
      
      Our best-fit parameters are listed in the last column of table~\ref{results}: our inferred chirp mass $M_c$ was correct to within $\Delta M_c/M_c  < 10^{-3}$, our inferred symmetric mass ratio $\eta$ to within $\Delta \eta \approx 4\times 10^{-3}$, and the error in our coalescence time was $\Delta t_c \approx 45$ s, corresponding to approximately $0.05$ GW periods just before the plunge.
Nevertheless, it is clear from our estimates for the \emph{other} parameters that, instead of converging on a neighborhood of the true maximum, our MCMC code locked onto a high but secondary maximum of the posterior probability distribution. 
Our inferred sky position is almost at the antipodes of the actual location (i.e., our ecliptic latitude is  approximately the negative of the true value, and our ecliptic longitude is off by nearly $\pi$).
This was not due to a mismatch of conventions or a bug in our code; 
rather, it reflects the above-mentioned degeneracy between antipodal sky 
locations (the degeneracy becomes perfect in the low-frequency limit).  
The four parameters $(D, \iota, \psi, \varphi_0)$ that determine the 
overall complex amplitudes of the GW polarizations $h_+$ and 
$h_{\times}$ were also off by factors of order one, except for our 
overall phase $\varphi_0$, which was correct to within $0.004$ radians 
(modulo $\pi$). 
\begin{table}
\caption{True values and estimates from three steps for the challenge 
parameters. In stages 1 and 2 estimates were made only for parameters 
$M_c$ and $\eta$ (and therefore $m_1$ and $m_2$) and 
$t_c$.\label{results}} 
\begin{indented}
\lineup
\item[]\begin{tabular}{l@{\quad}l@{\quad}l@{\quad}l@{\quad}l@{\quad}l}
\br
Parameter & Unit & True value & Stage 1 & Stage 2 & Stage 3 \\
\mr
$M_c$ & $10^6\ M_\odot$ & 1.2086 & 1.208 & 1.2108 & \01.2077 \\
$\eta$ & & 0.160 & 0.17 & 0.163 & \00.156 \\
$m_1$ & $10^6\ M_\odot$ & 2.8972 & 2.74 & 2.8536 & \02.9652 \\
$m_2$ & $10^6\ M_\odot$ & 0.7270 & 0.76 & 0.7381 & \00.7130 \\
$t_c$ & $10^7\ {\rm s}$ & 1.3374027 & 1.3372 & 1.3374149 & \01.3374072 \\
Ecl.\ Lat.\ $\theta$ & rad & \-0.492 & -- & -- & \00.536 \\
Ecl.\ Long.\ $\phi$ & rad & 0.866 & -- & -- & \04.039 \\
Pol.\ Angle $\psi$ & rad & 3.234 & -- &	-- & \05.886 \\
Init.\ Phase $\varphi_0$ & rad & 3.527 & -- & -- & \00.233 \\
Distance $D$ & $10^9$ pc & 8.000 & -- & -- & 16.811 \\
Incl.\ Angle $\iota$ & rad & 1.944 & -- & -- & \00.617 \\
\br
\end{tabular}
\end{indented}
\end{table}


It is also instructive (and reassuring) to contemplate the performance of the first two stages of our search. Stage 1 returned $M_c$ with a fractional error $\Delta M_c/M_c  <
10^{-3}$,  $\eta$ to within $\sim 6\%$, and $t_c$ to within $\sim 2\times 10^3$ s.  After stage 2, the estimated $M_c$ was in fact slightly worse, but the errors in $\eta$ and
$t_c$ were significantly reduced, to $\Delta \eta \approx 0.003$ and $\Delta t_c \approx 120$ s. This gratifying level of accuracy indicates that the coarser stages 1 and 2 were
indeed accomplishing the job required of them. 

\section{Future Directions}
\label{s:future}

   As explained above, the most obvious improvement to our pipeline will be to recast the MCMC stage so that it maximizes the ${\mathcal F}$-statistic on the 5-dimensional space
$(M_c,\eta,t_c,\theta,\phi)$, reducing the search-space dimensionality by three. In addition, we will extend our grid search to handle the case where the merger occurs
\emph{after} the end of the dataset (we did not compete on dataset 1.2.2 because our current grid search could not handle such mergers). This generalization should be fairly
straightforward.  

In the second round of Challenges (see the proceeding by Arnaud and colleagues in this volume~\cite{Arnaud:2007yj}), dataset 2.2 contains signals from an entire Galaxy's worth of
white-dwarf binaries, four to six SMBH binary inspirals (the exact number is not specified) with SNRs ranging from $\sim 10$ to $\sim 2000$, and five EMRIs.   Our plan is to first run
our pipeline as a standalone search for the SMBH binaries, and then to join forces with Crowder and Cornish's WD binary search~\cite{Crowder:2006eu} to iteratively improve the fits provided
by the two searches.  Beyond that, we plan to extend the SMBH binaries search to include: 1) merger and ringdown waveforms; 2) spin-precession effects; and 3) the effects of nonzero eccentricity. For the first two items, we intend to make use of the technology already developed by the ground-based GW community.  For instance, Buonanno, Chen, and Vallisneri~\cite{Buonanno:2002fy} have shown how
searches for binaries of spinning BHs can be made considerably more efficient by dividing the parameters into intrinsic (such as the masses) and extrinsic (such as the 
orientation of the orbital plane at a fiducial time), and optimizing over the extrinsic parameters semi-analytically.  (This can be viewed as a generalization to spinning binaries of the 
${\mathcal F}$-statistic analysis mentioned above.)
We shall endeavour to generalize this strategy to LISA
searches for SMBH binaries.

\ack
JC's, CC's and MV's work was carried out at the Jet Propulsion Laboratory,
California Institute of Technology, under contract to the National Aeronautics
and Space Administration. IM would like to thank the Brinson Foundation, NASA
grant NNG04GK98G and NSF grant PHY-0601459 for financial support. MV is
grateful for support from the Human Resources Development Fund program at JPL.
DB acknowledges funding from NSF grant PHY-0601459 and the LIGO Laboratory.
LIGO was constructed by the California Institute of Technology and
Massachusetts Institute of Technology with funding from the National Science
Foundation and operates under cooperative agreement PHY-0107417. This paper
carries LIGO Document Number LIGO-P070018-00-Z.

\vspace{-6pt}
\section*{References}
\vspace{-6pt}

\bibliographystyle{iopart-num}
\bibliography{MLDCproc}

\end{document}